entire latex paper,  no figures

\documentstyle[prl,aps]{revtex}
\begin{document}
\title{Color filamentation in ultrarelativistic \\ heavy-ion collisions}

\author{Stanis\l aw Mr\' owczy\' nski\footnote{Electronic address:
mrow@fuw.edu.pl}}

\address{High-Energy Department, So\l tan Institute for Nuclear Studies,\\
ul. Ho\.za 69, PL - 00-681 Warsaw, Poland}

\date{5-th June 1996}

\maketitle

\begin{abstract}

We study color fluctuations in the quark-gluon plasma produced at the early 
stage of nucleus-nucleus collision at RHIC or LHC. The fluctuating 
color current, which flows along the beam, can be very {\it large} due to 
the strong anisotropy of the parton momentum distribution. A specific 
fluctuation, which splits the parton system into the current filaments 
parallel to the beam direction, is argued to grow exponentially. The physical 
mechanism responsible for the phenomenon, which is known as a filamentation 
instability, is discussed.

\end{abstract}


In the near future the nucleus-nucleus collisions will be studied 
experimentally at the accelerators of a new generation: Relativistic 
Heavy-Ion Collider (RHIC) at Brookhaven and Large Hadron Collider (LHC) 
at CERN. The collision energy will be larger by one or even several orders 
of magnitude than that one of the currently operating machines.  A copious 
production of partons, mainly gluons, due to hard and semihard processes 
is expected in the heavy-ion collisions at this new energy domain 
\cite{Gei95}. Thus, one deals with the many-parton system at the early
stage of the collision. The system is on average locally colorless
but random fluctuations can break the neutrality. Since the system is 
initially far from equilibrium the color fluctuations can noticeably
influence its evolution.

In our previous papers \cite{Mro93} we have studied the plasma oscillations 
in the nonequilibrium quark-gluon system produced in ultrarelativistic 
heavy-ion collisions. We have argued that due to the strong anisotropy of 
the parton momentum distribution there are unstable modes which exponentially 
grow in time. Here we show that the fluctuations which initiate
these modes are {\it large}, much larger than in the equilibrium plasma. 
We also discuss the physical mechanism responsible for the growth of the 
fluctuation which splits the parton system into the color current filaments 
parallel to the beam direction.  For completeness we recapitulate at
the end of this letter the results from our papers \cite{Mro93}, which show 
the existence of the unstable modes.

The distribution functions of quarks $Q_{ij}(t,{\bf x},{\bf p})$, antiquarks 
$\bar Q_{ij}(t,{\bf x},{\bf p})$, and gluons $G_{ab}(t,{\bf x},{\bf p})$ 
with $i,j = 1, 2, 3$ and $a,b = 1, 2, ... , 8$ are \cite{Elz89,Mro89} matrices 
in the color space;  $t$, ${\bf x}$ and ${\bf p}$ denote the time, position, 
and the momentum. The color current expressed through these functions reads
\cite{Elz89,Mro89}
\begin{eqnarray*}
j^{\mu}_a(t,{\bf x}) = {1 \over 2} \,g \int {d^3p \over (2\pi )^3} \; 
{p^{\mu} \over E_p} \, 
\bigg(\tau^a_{ji}\Big( Q_{ij}(t,{\bf x},{\bf p}) 
- \bar Q_{ij}(t,{\bf x},{\bf p}) \Big)
+ i f^{abc} G_{bc}(t,{\bf x},{\bf p}) \bigg) \;, 
\end{eqnarray*}
where $g$ is the QCD coupling constant, $\tau^a$ is the $SU(3)$ group
generator, $f^{abc}$ the respective structure constant, and 
$p^{\mu} = (E_p, {\bf p})$ is the parton four-momentum  with 
$E_p = \vert \bf p \vert$ being the energy of the massless quark 
or gluon. 

We assume that the quark-gluon plasma is on average locally colorless, 
homogeneous, and stationary. Thus, the distribution functions averaged over 
ensemble are of the form
\begin{eqnarray*}
\langle Q_{ij}(t,{\bf x},{\bf p}) \rangle = 
\delta_{ij} n({\bf p}) \;,\;\;\;\;\;
\langle \bar Q_{ij}(t,{\bf x},{\bf p}) \rangle = 
\delta_{ij} \bar n({\bf p}) \;,\;\;\;\;\;
\langle G_{ij}(t,{\bf x},{\bf p}) \rangle = 
\delta_{ij} n_g({\bf p}) \;,
\end{eqnarray*} 
which give the zero average color current.

We find the fluctuations of the color current generalizing a well-known
formula for the electric current \cite{Akh75}. For a system of 
noninteracting quarks and gluons we have in the classical limit the 
following expression
\begin{eqnarray}\label{cur-cor-x}
M^{\mu \nu}_{ab} (t,{\bf x}) \buildrel \rm def \over = 
\langle j^{\mu}_a (t_1,{\bf x}_1) j^{\nu}_b (t_2,{\bf x}_2) \rangle 
= {1 \over 8} \,g^2\; \delta^{ab} 
\int {d^3p \over (2\pi )^3} \; {p^{\mu} p^{\nu} \over E_p^2} \;
f({\bf p}) \; \delta^{(3)} ({\bf x} -{\bf v} t)  \;,
\end{eqnarray}
where the effective parton distribution function $f({\bf p})$ equals 
$n({\bf p}) + \bar n({\bf p})  +  6 n_g({\bf p})$, 
$(t,{\bf x}) \equiv (t_2-t_1,{\bf x}_2-{\bf x}_1)$, and 
${\bf v} = {\bf p}/E_p$ is the parton velocity. Due to the average space-time 
homogeneity the correlation tensor depends only on the difference 
$(t_2-t_1,{\bf x}_2-{\bf x}_1)$. 

The physical meaning of the formula (\ref{cur-cor-x}) is  
transparent. The space-time points $(t_1,{\bf x}_1)$ and $(t_2,{\bf x}_2)$ 
are correlated in the system of noninteracting particles if  
the particles fly from $(t_1,{\bf x}_1)$ to $(t_2,{\bf x}_2)$. 
Therefore the delta  $\delta^{(3)} ({\bf x} - {\bf v} t)$ is present 
in the formula (\ref{cur-cor-x}). The momentum integral of the
distribution function simply represents the summation over particles.

One finds the Fourier spectrum of the fluctuations from 
eq.~(\ref{cur-cor-x}) as
\begin{equation}\label{cur-cor-k}
M^{\mu \nu}_{ab} (\omega ,{\bf k}) = {1 \over 8} \,g^2\; \delta^{ab} 
\int {d^3p \over (2\pi )^3} \; 
{p^{\mu} p^{\nu} \over E_p^2} \; f({\bf p})  \;
2\pi \delta (\omega -{\bf kv}) \;,
\end{equation}
where $\omega$ is the frequency and ${\bf k}$ the wave-vector.

We model the parton momentum distribution at the early stage of 
ultrarelativistic heavy-ion collision in two ways:
\begin{equation}\label{f-flat-y}
f({\bf p}) = {1 \over 2Y} 
\Theta(Y - y) \Theta(Y + y) \; h(p_{\bot}) \;
{1 \over p_{\bot} \, {\rm ch}y}\;, 
\end{equation}
and
\begin{equation}\label{f-flat-pl}
f({\bf p}) = {1 \over 2{\cal P}} 
\Theta({\cal P} - p_{\parallel}) 
\Theta({\cal P} + p_{\parallel}) \; 
h(p_{\bot}) \;, 
\end{equation}
where $y$, $p_{\parallel}$, $p_{\bot}$, $y$, and $\phi$ denote the parton 
rapidity, the longitudinal and transverse momenta, and the azimuthal angle, 
respectively. The parton momentum distribution (\ref{f-flat-y}) corresponds
to the rapidity distribution which is flat in the interval $(-Y,Y)$.
The distribution (\ref{f-flat-pl}) is flat for the longitudinal 
momentum $-{\cal P} < p_{\parallel} < {\cal P}$. We do not specify the 
transverse momentum distribution $h(p_{\bot})$, which is assumed to be 
of the same shape for quarks and gluons, because it is sufficient for 
our considerations to demand that the distributions 
(\ref{f-flat-y}, \ref{f-flat-pl}), are strongly elongated along the 
$z-$axis i.e. $e^Y \gg 1$ and 
$\langle p_{\parallel} \rangle \gg \langle p_{\bot} \rangle$. 

Due to the symmetry $f({\bf p}) = f(-{\bf p})$ of the distributions
(\ref{f-flat-y},\ref{f-flat-pl}), the correlation tensor $M^{\mu \nu}$ 
is diagonal i.e. $M^{\mu \nu}= 0$ for $\mu \not= \nu$.  Since the average 
parton longitudinal momentum  is much bigger than the transverse one,
it obviously follows from eq.~(\ref{cur-cor-k}) that the largest 
fluctuating current appears along the $z-$axis. Therefore, we discuss 
the $M^{zz}$ component of the correlation tensor. $M^{zz}(\omega, {\bf k})$ 
depends on the ${\bf k}-$vector orientation and there are two generic cases: 
${\bf k} = (k_x,0,0)$ and ${\bf k} = (0,0,k_z)$. The inspection of 
eq.~(\ref{cur-cor-k}) shows that the fluctuations with ${\bf k} = (k_x,0,0)$ 
are much larger than those with ${\bf k} = (0,0,k_z)$.
Thus, we compute $M^{zz}(\omega, k_x)$. 

Substituting the distributions (\ref{f-flat-y},\ref{f-flat-pl}) 
into (\ref{cur-cor-k}) one finds after azimuthal integration
\begin{eqnarray}\label{cor-zzx-flat-y}
M^{zz}_{ab}(\omega, k_x) = \delta^{ab}
\; {g^2 \over 32\pi^2\, Y} \int^Y_{-Y} dy \; \int_0^{\infty} dp_{\bot}
\; h(p_{\bot}) \, p_{\bot} \;
{{\rm sh}^2y \over {\rm ch}y } \;
{\Theta (k_x^2 - \omega^2 \,{\rm ch}^2y ) \over  
\sqrt{ k_x^2  - \omega^2  \, {\rm ch}^2y } } \;,
\end{eqnarray}
\begin{eqnarray}\label{cor-zzx-flat-pl}
M^{zz}_{ab}(\omega, k_x) = \delta^{ab} 
\;{g^2 \over 32\pi^2{\cal P} } \;
\int^{\cal P}_{\cal -P} dp_{\parallel} \; 
\int_0^{\infty} dp_{\bot} \;
h(p_{\bot}) \; { p_{\bot} p_{\parallel}^2 \over E_p } \;
{\Theta (k_x^2 p_{\bot}^2 - \omega^2 E_p^2) \over 
\sqrt{k_x^2 p_{\bot}^2 - \omega^2 E_p^2}} \;.
\end{eqnarray}

One observes that the integrals from 
eqs.~(\ref{cor-zzx-flat-y},\ref{cor-zzx-flat-pl}) reach
the maximal values for $\omega^2 \ll k_x^2$. Therefore, we compue 
$M^{zz}_{ab}$ for $\omega= 0$. Keeping in mind that 
$e^Y \gg 1$ and $\langle p_{\parallel} \rangle \gg \langle p_{\bot} \rangle$ 
we get the following approximate expressions for the flat $y-$ and 
$p_{\parallel}-$distributions:  
\begin{eqnarray}\label{cor-zzx-0-flat-y}
M^{zz}_{ab}(\omega=0, k_x) = 
{1 \over 8} \,g^2\: \delta^{ab} \; {e^Y \over Y} \;
{\langle \rho \rangle  \over \vert k_x \vert } \;,
\end{eqnarray}
\begin{eqnarray}\label{cor-zzx-0-flat-pl}
M^{zz}_{ab}(\omega=0, k_x) =  
{1 \over 8} \,g^2\: \delta^{ab} \; 
{{\cal P} \over \langle p_{\bot} \rangle } \;
{\langle \rho \rangle  \over \vert k_x \vert } \;,
\end{eqnarray}
where $\langle \rho \rangle$ is the effective parton density given as
\begin{eqnarray*}
\langle \rho \rangle \equiv \int {d^3p \over (2\pi )^3} \; f({\bf p})  
= {1 \over 4\pi^2} \int_0^{\infty} dp_{\bot}p_{\bot} \; h(p_{\bot}) 
= {1\over 3} \; \langle \rho \rangle_{q\bar q} +
{3 \over 4} \; \langle \rho \rangle_g \;,
\end{eqnarray*}
with $\langle \rho \rangle_{q\bar q}$ denoting the average density of 
quarks and antiquarks, and $\langle \rho \rangle_g$ that of gluons.
For the flat $p_{\parallel}-$case we have also used the approximate 
equality
\begin{eqnarray*}
\int_0^{\infty} dp_{\bot} \; h(p_{\bot}) =
{1 \over \langle p_{\bot} \rangle }
\int_0^{\infty} dp_{\bot}p_{\bot} \; h(p_{\bot}) 
\end{eqnarray*}
to get the expression (\ref{cor-zzx-0-flat-pl})

It is instructive to compare the results (\ref{cor-zzx-0-flat-y},
\ref{cor-zzx-0-flat-pl}) with the analogous one for the equilibrium 
plasma which is
\begin{eqnarray*}
M^{zz}_{ab}(\omega=0, k_x) = 
{\pi \over 16} \,g^2\; \delta^{ab}\;
{ \langle \rho \rangle \over \vert k_x \vert } \;. 
\end{eqnarray*}
One sees that {\it the current fluctuations in the anisotropic plasma 
are amplified by the large factor} which is $e^Y/Y$ or
${\cal P} /\langle p_{\bot} \rangle$.

Let us now discuss how the fluctuation, which contributes to 
$M^{zz}_{ab}(\omega=0, k_x)$, evolves in time. The form of 
the fluctuating current is
\begin{eqnarray}\label{flu-cur}
{\bf j}_a(x) = j_a \: \hat {\bf e}_z \: {\rm cos}(k_x x) \;,
\end{eqnarray}
where $\hat {\bf e}_z$ is the unit vector in the $z-$direction.
Thus, there are current filaments of the thickness $\pi /\vert k_x\vert$ 
with the current flowing in the opposite directions in the neighboring 
filaments.

In the limit of weak fields the chromodynamics can be approximately
treated as an eight-fold electrodynamics. Consequently, the magnetic 
field generated by the current (\ref{flu-cur}) is given as
\begin{eqnarray*}
{\bf B}_a(x) = {j_a \over k_x} \: \hat {\bf e}_y \: {\rm sin}(k_x x) \;.
\end{eqnarray*}
The Lorentz force acting on the partons, which fly along the beam, equals
\begin{eqnarray*}
{\bf F}(x) = q_a \: {\bf v} \times {\bf B}_a(x) = 
- q_a \: v_z \: {j_a \over k_x} \: \hat {\bf e}_x \: {\rm sin}(k_x x) \;,
\end{eqnarray*}
where $q_a$ is the color charge. One observes, see Fig.~1, that the force 
distributes the partons in such a way that those, which positively 
contribute to the current in a given filament, are focused to the 
filament center while those, which negatively contribute, are moved 
to the neighboring one. Thus, the initial current is growing. 

The mechanism described here is well-known in the plasma physics 
\cite{Che84} and it leads to the so-called filamentation or Weibel 
instability \cite{Wei59}.  In the context of the quark-gluon plasma
the phenomenon has been first discussed in the system of two interpenetrating 
parton beams \cite{Mro88}. Such a system however seems to be completely 
unrealistic from the experimental point of view. Then it has been argued
\cite{Mro93} that the filamentation instability can occur under the 
conditions which will be realized in heavy-ion collisions at RHIC and LHC.
Let us briefly recapitulate these considerations which confirm the 
qualitative arguments presented above. 

The spectrum of plasma modes initiated by the color fluctuations 
is determined by the dispersion equation which for the anisotropic 
plasma is 
\begin{eqnarray*}
{\rm det} \vert {\bf k}^2 \delta ^{ij} - k^i k^j - 
\omega ^2  \epsilon ^{ij}(\omega, {\bf k}) \vert = 0 \;, 
\;\;\;\;\; i,j = x, y, z 
\end{eqnarray*}
where the chromodielectric tensor $\epsilon ^{ij}$ is 
\begin{eqnarray*}
\epsilon ^{ij} (\omega, {\bf k}) = \delta ^{ij} + 
{ g^2 \over 2\omega} \int {d^3 p \over (2\pi )^3}
{ v^i \over \omega - {\bf k v} + i0^+} 
{\partial f({\bf p}) \over \partial p^l} 
\Bigg[ \Big( 1 - {{\bf k v} \over \omega} \Big) \delta ^{lj}
+ {k^l v^j \over \omega} \Bigg] \;.
\end{eqnarray*}

For the specific color fluctuation with wave vector along the 
$x-$axis and the electric field parallel to the $z-$direction, 
the dispersion equation simplifies to 
\begin{eqnarray}\label{disp-eq}
H(\omega) \equiv k_x^2 - \omega ^2 \epsilon ^{zz}(\omega, k_x) = 0 \;.
\end{eqnarray}
The so-called Penrose criterion states that {\it the dispersion equation 
$H(\omega )=0$ has unstable solutions if} $H(\omega = 0) < 0$. We have shown 
that under the reasonable assumptions concerning the form of the transverse 
momentum distribution $h(p_{\bot})$, the criterion is indeed satisfied. Then, 
we have solved approximately the dispersion equation (\ref{disp-eq}) and found 
the unstable mode with the pure imaginary frequency. The estimated 
characteristic time of the instability development, which appears as short as 
$0.3 - 0.4 \;\; {\rm fm}/c$, is significantly smaller than the life time
of the plasma created in the nuclear collision. Therefore, the instability
can indeed develop and play a significant role in the system dynamics.

One asks whether the color instabilities are detectable in ultrarelativistic
heavy-ion collisions. The answer seems to be positive because the 
occurrence of the filamentation breaks the azimuthal symmetry of the system 
and hopefully will be visible in the final state. The azimuthal orientation 
of the wave vector will change from one collision to another while the 
instability growth will lead to the energy transport along this vector 
(the Poynting vector points in this direction). Consequently, one expects 
significant variation of the transverse energy as a function of the azimuthal
angle. This expectation is qualitatively different than that based on the 
parton cascade simulations \cite{Gei95}, where the fluctuations are strongly 
damped due to the large number of uncorrelated partons. Due to the collective 
character of the filamentation instability the azimuthal symmetry will be 
presumably broken by a flow of large number of particles with relatively 
small transverse momenta. The jets produced in hard parton-parton 
interactions also break the azimuthal symmetry. In this case however
the symmetry is broken due to a few particles with large transverse momentum.
The problem obviously needs further studies but it seems
that the event-by-event analysis of the nuclear collision give a chance 
to observe the color instabilities in the experiments planed at RHIC
and LHC. 

This work was partially supported by the Polish Committee of Scientific 
Research under Grant No. 2 P03B 195 09.

\vspace{1cm}
\subsection*{Figure captions}
\noindent
{\bf FIG.~1.} The mechanism of filamentation. The phenomenon is, for
simplicity, considered in terms of the electrodynamics. The fluctuating 
current generates the magnetic field acting on the positively charged 
particles which in turn contribute to the current (see text). 
$\otimes$ and $\odot$ denote the parallel and, respectively, antiparallel 
orientation of the magnetic field with respect to the $y-$axis.

\end{document}